\documentclass[twocolumn]{revtex4}
\usepackage{amssymb,epsfig,amscd}
\usepackage{graphicx}
\usepackage{dcolumn}
\usepackage{bm}

\usepackage{epsfig}

\newcommand{\dirac}{\partial\llap{$\diagup$\kern-2pt}}

\def\be{\begin{equation}} 
\def\ee{\end{equation}}
\def\bq{\begin{eqnarray}} 
\def\eq{\end{eqnarray}}

\begin{document}

\title{A new possible quark-hadron mixed phase in protoneutron stars}
\author{G.~Pagliara, M.~Hempel, J.~Schaffner-Bielich}

\affiliation{Institut f\"{u}r Theoretische Physik, Ruprecht-Karls-Universit\"at,
   Philosophenweg 16,  D-69120, Heidelberg, Germany}

\begin{abstract} 
The phase transition from hadronic matter to quark matter at high
density might be a strong first order phase transition in presence of
a large surface tension between the two phases. While this implies a
constant-pressure mixed phase for cold and catalyzed matter this is
not the case for the hot and lepton rich matter formed in a
protoneutron star. We show that it is possible to obtain a mixed phase
with non-constant pressure by considering the global conservation of
lepton number during the stage of neutrino trapping. In turn, it
allows for the appearance of a new kind of mixed phase as long as
neutrinos are trapped and its gradual disappearance during
deleptonization. This new mixed phase, being composed by two electric
neutral phases, does not develop a Coulomb
lattice and it is formed only by spherical structures, drops and
bubbles, which can have macroscopic sizes. The disappearance of the
mixed phase at the end of deleptonization might lead to a delayed
collapse of the star into a more compact configuration containing a core of pure
quark phase. In this scenario, a significant emission of neutrinos
and, possibly, gravitational waves are expected.
\end{abstract}

\maketitle

The possibility of the formation of a quark hadron mixed phase in neutron
stars, especially during the first stages after their birth, has been
widely discussed in the literature
and interesting associated signatures were proposed \cite{Prakash:1995uw,Drago:1997tn,Steiner:2000bi,Panda:2003dh,Nicotra:2006eg}.
In Ref.~\cite{Pons:2001ar} a delayed formation of the quark phase was
found, after the deleptonization of the star, which can trigger the collapse 
of the protoneutron star to a black hole. An observation of a
supernova neutrino signal with an abrupt cessation of the signal would
be a clear confirmation of this scenario. Recently, a similar result was
obtained in Ref.~\cite{Nakazato:2008su} where a supernova simulation
for a 100 $M_\odot$ progenitor star was performed. The appearance of
quark matter reduces the delay between the core bounce and the
collapse to a black hole. In both above mentioned paper
the MIT bag model was used
to describe quark matter and a large value of the
bag constant was adopted resulting in a large critical density
for the appearance of quark matter. By assuming a small bag constant
value, which still allows to obtain maximum masses for hybrid stars
compatible with observations, it was found in Ref.~\cite{Sagert:2008ka} 
that quark matter can form already during the early post bounce phase
of a core-collapse supernova. Interestingly, the phase transition leads to the formation of a shock
wave propagating outwards which provides, in some cases, the explosion of the supernova even in
spherical symmetry. Moreover a new neutrino burst is emitted with a typical time delay with
respect to the first neutronization burst of a few hundreds of
milliseconds.
In all these studies the Gibbs construction \cite{Glendenning:1992vb}
is adopted to describe the
quark hadron mixed phase but finite size effects are not included i.e. one assumes that the surface tension $\sigma$ between the hadronic
phase and the quark phase is vanishingly small. Presently, the value of
$\sigma$ is~not~known and the possibility that it has a large value, of say 
$\sigma \sim 100$ MeV/fm$^2$, cannot be excluded \cite{Alford:2001zr}. If
this is indeed the case, the mixed phase equation of state must be
computed including the effects from the surface energy $\epsilon_S$ and the Coulomb energy
$\epsilon_C$ of the finite geometrical structures, which are dubbed the ``pasta phases'' \cite{Heiselberg:1992dx,Glendenning:1995rd}.  
The optimal size and shape of a structure at a fixed density is determined by
the minimization of the energy which gives the well known relation: $\epsilon_S=2\epsilon_C
$. It was pointed out in Refs.~\cite{Voskresensky:2002hu,Endo:2005zt}
that also the effect of charge screening and the rearrangement
of charged particles in presence of the Coulomb interactions
must be taken into account for a realistic description of the mixed phase.
In fact local charge neutrality is almost recovered at lengths much larger than the Debye screening length. 
Therefore the mixed phase window shrinks
considerably and it approaches the constant-pressure Maxwell
construction \cite{Voskresensky:2002hu,Endo:2005zt}. This effect in turn would imply the
absence of the mixed phase in cold and deleptonized hybrid stars. 
A calculation of finite size and charge screening effects in the mixed phase
for protoneutron star matter, which has fixed entropy per baryon and
fixed lepton fraction, has not yet been performed. Here we want to
consider this possibility but instead of solving the system of
Poisson equations, which is necessary to obtain the density profiles 
of charged particles, we assume local charge neutrality for the two phases as in the
Maxwell construction (this simple treatment is justified if $\sigma$ is sufficiently large). 
Consequently, we introduce two distinct charge
chemical potentials: $\mu_C^h$ for the hadronic phase and $\mu_C^q$ for
the quark phase and one baryonic chemical potential $\mu_B$ corresponding
to the global conservation of the baryon number.
In a newly born neutron star however, besides the conservation of
baryon number also the lepton number is conserved as long as neutrinos are trapped, i.e.
during the first $\sim 10$ seconds of the life of the star.
Neutrino trapping allows to introduce a chemical potential
$\mu_L$, associated with the additional conserved leptonic charge, which
coincides with the chemical potential of neutrinos $\mu_L=\mu_\nu$ (see \cite{Matthias} for a general discussion on mixed phases). 
Let us consider the interface between the two phases: 
a charge separated interface is formed with a size of the order of the Debye screening length, $\sim 10$ fm,
with a layer of positively charged, electron depleted, hadronic matter on one side
and a layer of quark matter with an excess of the electron on the other side (as discussed in \cite{Alford:2001zr} for the CFL phase). The interface is 
stabilized by the resulting electric field.
Notice that neutrinos, being not affected by the electric field,
can freely stream across the interface.
Consequently, lepton number is conserved only globally. This additional
globally conserved quantum number has similar effects as the global charge neutrality condition
adopted to model the phase transition 
for vanishing values of the surface tension.
 \begin{figure}
    \begin{centering}
\epsfig{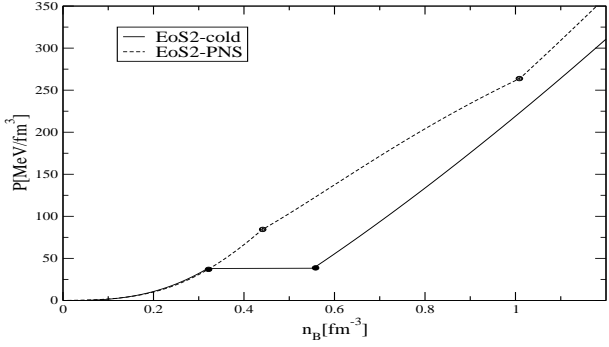}
    \caption{The equations of state are shown for the case of EoS2 for protoneutron star matter (dashed line) and 
for cold and catalyzed star matter (solid line). The dots indicate the onset and the end of the mixed phase in both cases.
\label{fig:eos} }
   \end{centering}
   \end{figure} 
We consider here the ``standard'' conditions of a newly born 
neutron star \cite{Steiner:2000bi}: the matter has a fixed lepton fraction $Y_L=(n_e+n_{\nu})/n_B=0.4$ and 
fixed entropy per baryon $S/N_B=1$ where $n_e$, $ n_{\nu}$ and $n_B$ are 
the electron, neutrino and baryon number densities and $S$ is the entropy.
The conditions of chemical-equilibrium in the two phases read:
\begin{eqnarray}
\mu_n=\mu_B,\,\,\mu_p=\mu_B+\mu^h_C\\
\mu_u=(\mu_B+2\mu^q_C)/3,\,\, \mu_d=(\mu_B-\mu^q_C)/3,\,\, \mu_s=\mu_d   \\ 
\mu^h_e=\mu_L-\mu^h_C,\,\, \mu^q_e=\mu_L-\mu^q_C,\,\,\mu^h_\nu=\mu^q_\nu=\mu_L 
\end{eqnarray}
where $\mu_i (i=n,p,e,\nu,u,d,s)$ are the chemical potentials of neutrons,
protons, electrons, neutrinos and up down and strange quarks, respectively. The Gibbs
conditions for mechanical and thermal equilibrium read:
\begin{eqnarray}
P^h(\mu_B,\mu^h_C,\mu_L,T) = P^q(\mu_B,\mu^q_C,\mu_L,T)\\
n^h_C-n^h_e=0 \,\, ,  n^q_C-n^q_e=0 \\
(1-\chi)(n^h_e+n_{\nu})+ \chi(n^q_e+n_{\nu})= Y_L n_B\\
(1-\chi)s^h+\chi s^q= S/Nn_B
\end{eqnarray}
where $P^{\alpha}$, $s^{\alpha}$ with $\alpha=h,q$ are the pressure and the entropy density in the
hadronic and quark phases, $\chi$ is the volume fraction of the quark
phase and $n^{\alpha}_C$ are the electric charge densities of hadrons and quarks. As
usual the baryon density is defined as follows:
$n_B=(1-\chi)n^h_B+\chi n^q_B $. Notice that the eqs. (5) impose
local charge neutrality of the two phases and eq. (6) global lepton
number conservation.
\begin{figure}
    \begin{centering}
\epsfig{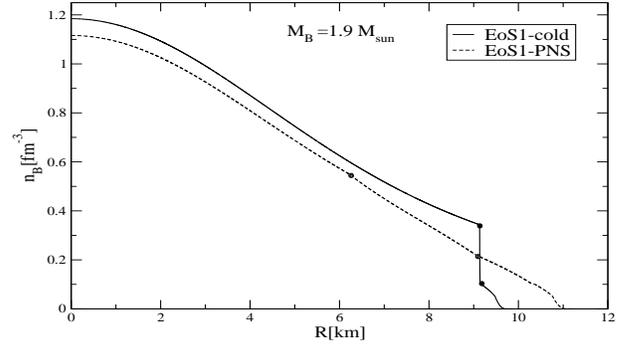}
    \caption{Density profiles for a star with a baryon mass of $1.9 M_{\odot}$
for the protoneutron star stage and the cold configuration. The dots mark the onset and the end of the phase transition.
\label{fig:prof} }
   \end{centering}
   \end{figure} 
To calculate the equations of state of hadronic matter and quark
matter we adopt the relativistic mean field model with the
parametrization TM1 for the former \cite{Shen:1998gq} and the MIT bag
model including perturbative corrections for the latter \cite{Fraga:2001id,Alford:2004pf}. 
We set the masses of up and down quarks to zero and the
mass of the strange quark to $100$ MeV. 
We fix the constant which simulates the QCD perturbative corrections $c=0.2$,
and we select two values of the effective bag constant $B_{\rm eff}$ in order to have a critical
density for the phase transition in protoneutron star matter of $\sim
n_0$ and $\sim 3 n_0$ (where $n_0=0.16$ fm$^{-3}$ is the nuclear saturation density), corresponding to $B_{\rm eff}^{1/4}=155$~MeV and
$B_{\rm eff}^{1/4}=170$ MeV. 
The two equations of state are labeled as
EoS1 and EoS2 for the two choices of the effective bag constant. 
In Fig.~1 we show the equations of state for matter in a protoneutron
star (indicated with PNS) and for cold and catalyzed matter (indicated
with ``cold'').  The remarkable result is that within the mixed phase
constructed by solving eqs. (4)-(7), the pressure increases as a
function of the density and a large range of density is occupied by
the mixed phase. During deleptonization the pressure in the mixed
phase gradually flattens and finally for deleptonized and cold matter
one finds the usual result of a Maxwell construction with a constant
pressure from the onset to the end of the phase transition.
We use now the above presented equations of state to study the
structures of protoneutron stars and cold stars. 
In Fig.~2 we show the density profile for a protoneutron star and the corresponding
cold configuration (assuming total baryon number conservation
during the cooling and deleptonization of the newly born
star.)  The mixed phase, initially present in a $\sim 2$ km large layer of
the star, gradually shrinks during the deleptonization of the star and
finally disappears in the cold configuration. As a result, a sharp interface
separating hadronic matter from quark matter is obtained 
with a sizable jump of the baryon density.
 \begin{figure}
    \begin{centering}
\epsfig{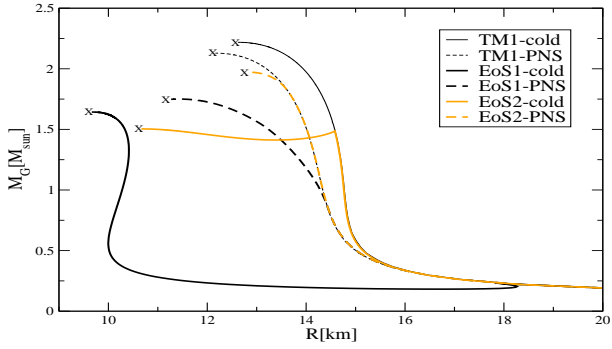}
    \caption{The mass-radius relations are shown for the different equations of state, EoS1, EoS2 and TM1 
and for protoneutron stars and cold stars configurations (colors online). The crosses correspond to the maximum mass configurations. 
\label{fig:mr} }
   \end{centering}
   \end{figure} 
In Fig.~3 we show the mass-radius relations for the different cases.
The black and grey (orange online) thick dashed lines correspond to hybrid
protoneutron stars (EoS1 and EoS2 respectively), the black and grey (orange online)
thick solid lines correspond to the cold configurations. Neutron stars mass-radius relations are also
shown for comparison (thin curves labeled with TM1).  The new mixed phase
appears in a protoneutron star because the pressure increases with the
density. Therefore, we obtain stellar configurations with a core of
pure quark matter, a layer of mixed phase and a layer/crust of
hadronic matter for the case of EoS1 and hybrid stars with only a core
of mixed phase in the case of EoS2. The mixed phase cannot appear
anymore in the star for cold and catalyzed matter because the pressure
is constant and only configurations with pure phases are obtained. As
discussed before, a sizable jump of the density occurs at the
interface separating the two pure phases which affects the stability of
the stars: at the onset of the phase transition the stars are
gravitationally unstable and only if a sizable volume of the star is
occupied by the quark phase the stars are stable.
The mass-radius relations in this case correspond to the so called
``third family'' solutions, see
\cite{1968PhRv..172.1325G,Schertler:2000xq,SchaffnerBielich:2002ki,Banik:2002kc}
for a detailed discussion of the properties of these stars.
We remark that the new mixed phase introduced here has quite different
properties with respect to the globally electric neutral
Gibbs mixed phase: since the two phases are locally charge neutral no
Coulomb lattice with charged finite structures of the two phases can
form. Instead an amorphous phase with only spherical pasta structures
is present (1-D and 2-D structures can indeed form only due to Coulomb
interactions). Moreover the charge neutral structures have
macroscopic sizes contrary to a globally electric neutral mixed phase
where the optimal size of the structures is
limited by the Coulomb energy. To reduce the total surface energy the
charge neutral structures start to merge and asymptotically a full separation of the
two phases will be obtained. Therefore in such a mixed phase
no coherent scattering of neutrinos with pasta structures
can take place \cite{Reddy:1999ad}, as the neutrino wavelength is much
smaller than the size of the structures. A detailed simulation of
neutrino transport within this new mixed phase would be extremely
interesting for the possible implications on the neutrino signal of
the changes of the structure of the star during deleptonization.
Also the motion and the interactions of the drops/bubbles within the mixed phase, in presence
of turbulence, might represent an interesting source of gravitational waves \cite{Megevand:2008mg}. 
\begin{figure}
    \begin{centering}
\epsfig{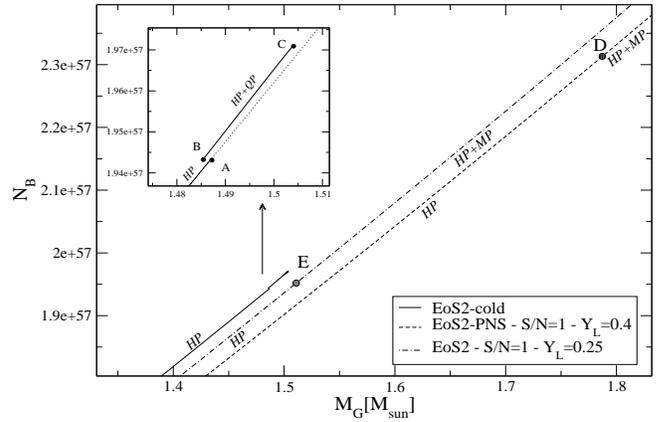}
    \caption{The total baryon number of compact stars as a function
of the gravitational mass for EoS2 at different stages of the temporal evolution:
protoneutron star (PNS, $Y_L$=0.4, $S/N=1$), an intermediate stage with $Y_L=0.25$, $S/N=1$
and the cold configurations (cold). In the insert we show a magnification of the third family branch (solid line)
and we also show the result for neutron stars (dotted line).
{\it HP}, {\it  MP} and {\it QP} denote the hadronic phase, the mixed phase and the quark phase respectively.
\label{fig:nbmg} }
   \end{centering}
   \end{figure} 
Concerning phenomenology, an interesting possibility is a delayed transition
of a protoneutron star in a third family star during/after
deleptonization, which is outlined in Fig.~4. The plot shows the baryon number of the stars as
a function of the gravitational mass for protoneutron stars (dashed
line) and cold stars (solid line). For the sake of discussion an intermediate configuration
is also included (dotted-dashed line) corresponding to partially deleptonized
matter with $Y_L=0.25$. The insert shows a magnification 
of the third family branch of cold hybrid stars. The letter A in the
plot denotes the configuration of a cold neutron star which is
unstable with respect to the collapse to a third family star, indicated
by the letter B, with the same baryon number. The energy released in
such a collapse (the difference between the gravitational masses
of the two configurations at fixed baryon number) is of the order of
$10^{51}$ ergs similar to values found in
Ref.~\cite{Mishustin:2002xe}. The letter C marks the maximum mass
of cold hybrid stars. The letter D stands for the configuration of a
protoneutron star for which the central density is equal to the
density of the onset of the phase transition
and a core of mixed phase is formed; the core of
mixed phase increases with the central density and therefore with the mass of the stars.
Stellar configurations with a baryon number lower than A are always
composed of purely hadronic matter during the evolution of the star.  
Since C is smaller than D, for stellar configurations having baryon number between the labels A and C
the corresponding protoneutron stars do not have quark matter in the core
(neither pure phase nor mixed phase) but during the deleptonization,
since the onset of the phase transition decreases, at a certain point a core
of mixed phase forms (e.g. at point E). As the
deleptonization proceeds the mixed phase gradually shrinks, a pure
quark phase core starts to form and finally for fully deleptonized
matter the mixed phase disappears and an hybrid star with pure phases
is obtained. Depending on the detailed dynamics of the formation of the
pure quark phase core and the disappearing of the mixed phase it is
possible that the evolution towards the final cold hybrid star
configuration, for stars having a baryon number close to A, proceeds through a gravitational collapse (similarly to
the transition from A to B). In that case the gravitational potential energy
is released in a short amount of time and a burst of neutrinos and
gamma rays can be produced as proposed in Ref.~\cite{Mishustin:2002xe}
for the collapse to third family stars. In such a fast dynamics also
gravitational waves might be emitted if non-radial modes are excited.
On the other hand it is also possible that 
the evolution of the star proceeds through hydrostatical equilibrium configurations, most probably for stars
having a baryon number close to C,
and the gravitational potential energy is released gradually. No strong signature
is expected in this case unless finite size effects
do play an important role for the nucleation of the new phase and the hadronic star can be 
in a metastable state before converting into a hybrid star \cite{Drago:2008tb}. Finally, stars having baryon number larger than C will develop a core of mixed phase
during deleptonization and will collapse to a black hole after the full 
deleptonization (similarly to the results of Ref.~\cite{Pons:2001ar}). 

In this paper we have shown that also in presence of a large value of
the surface tension between quark matter and hadronic matter it is
still possible to form a quark-hadron mixed phase in protoneutron
stars. Although the two phases are locally charge neutral, in protoneutron star matter the
existence of an additional globally
conserved charge, the total lepton number, allows to obtain a mixed
phase with non-constant pressure. The new mixed phase found here would
be present only during the first seconds of the life of the star and
it would gradually disappear as neutrinos become untrapped. The
possible effects on the temporal evolution of newly born hybrid stars
were also discussed in connection with the scenario of a delayed
transition of a neutron star to a third family star. We used here the
simple MIT bag model to compute the equation of state of quark matter
but our system of equations for the mixed phase has a general validity.
Nevertheless, it would be important to repeat the calculations by using others
models like for instance the NJL model \cite{Ruester:2005ib,Sandin:2007zr,Pagliara:2007ph}.  
Finally, the effects of the formation of this new mixed phase should be
investigated quantitatively in
supernova simulations and in calculations of neutrino transport
in protoneutron stars. 

The work of G.~P. is supported by the Alliance Program of the Helmholtz Association (HA216/EMMI).
M.~H. acknowledges support from the
Graduate Program for Hadron and Ion Research.
J.~S.~B. is supported by the German Research Foundation (DFG) through the Heidelberg Graduate School of Fundamental Physics. 
We thank Alessandro Drago, Irina Sagert and Sanjay Reddy for many fruitful discussions.
We also thank the CompStar program of the European Science Foundation.


\end{document}